%Paper: hep-th/9207030
%From: Thomas Larsson <tl@theophys.kth.se>
%Date: Thu, 9 Jul 92 17:05:32 EDT

% plain TeX

\def\Reals{{\bf R}}
\def\Ints{{\bf Z}}

\def\oj{{\bf g}}

\def\implies{\Longrightarrow}

\def\TT{{\bf T}}
\def\CC{{\bf C}}

\def\IX{\hbox{\raise.23ex\hbox{{$\scriptscriptstyle\mid$}}
\kern-0.62em\hbox{{$\times$}}}}

\def\d{\partial}
\def\ddot{{\kern-0.07em\cdot\kern-0.07em}}
\def\:{{\kern0.02em:\kern0.02em}}

\def\e{\hbox{e}}
\def\emx{\e^{m \ddot x}}
\def\enx{\e^{n \ddot x}}
\def\emnx{\e^{(m+n) \ddot x}}

\def\args#1#2{{{#1_1} \ldots {#1_{#2}}}}
\def\argsx#1#2#3{{{#1_1} \ldots #3 \ldots {#1_{#2}}}}

\def\nup{\args \nu p}

\def\tauq{\args \tau q}
\def\Ap{\args A p}
\def\Bq{\args B q}
\def\Aqp{\args A {{q-p-1}}}
\def\Bp{\args B p}

\def\pqefac{ {1 \over {(p+q+1)!}}}
\def\pqfac{ {1 \over {(p+q)!}}}
\def\pefac{ {1 \over {(p+1)!}}}
\def\qefac{ {1 \over {(q+1)!}}}

\def\xxx{\leftrightarrow}

\def\axb{a \xxx b}
\def\AxB{A \xxx B}

\def\phin{\tilde \phi}
\def\psin{\tilde \psi}
\def\thetan{\tilde \theta}

\def\en{\tilde e}

\def\dn{\tilde d}
\def\Dn{\tilde D}
\def\gamman{\tilde \gamma}

\def\proclaim#1{{\vskip 2mm \goodbreak \hbox{\bf #1} \vskip 1mm}}
\def\endproclaim{\vskip 2 mm}
\def\demo#1{{\it#1: }}
\def\enddemo{\vskip 2 mm}
\def\qed{\hfill{\vrule height 3mm width 2mm depth 0mm}}

\def\half{{1 \over 2}}

\def\bis{{\prime\prime}}
\def\mapleft#1{\hbox{$\smash{\mathop{\longleftarrow}\limits^{#1}}$}}
\def\mapright#1{\hbox{$\smash{\mathop{\longrightarrow}\limits^{#1}}$}}

\def\ref#1{{${}^{#1}$}}
\def\header#1{{\bf #1} \nobreak}

\def\endchapter{{\vskip 15 mm} \goodbreak}
\def\newpage{{\vfill \eject}}
\baselineskip = 12pt

\overfullrule = 0pt
\magnification = 1200

\centerline{\bf Vect(N) conformal fields and their exterior derivatives}

\vglue 30pt

T. A. Larsson\footnote*{Supported by the Swedish
Natural Science Research Council (NFR)}

Department of Theoretical Physics,

Royal Institute of Technology

100 44 Stockholm, Sweden

email: tl@theophys.kth.se

\vglue 60pt
(February 1992)
\vglue 60pt

\header{Abstract}

Conformal fields are a recently discovered class of representations of
the algebra of vector fields in $N$ dimensions. Invariant first-order
differential operators (exterior derivatives) for conformal fields are
constructed.

\vglue 1 in


\newpage

\header{1. Introduction}

In order to describe physical quantities, a coordinate system has to
be introduced, but physics itself should not depend on how this is
done. Therefore any sensible object must transform as a representation
of the group of diffeomorphisms (coordinate transformations) in the
relevant space, and conversely the representation theory of the
diffeomorphism group amounts to a classification of inequivalent
meaningful objects.

A sensible starting point is the Lie algebra of the diffeomorphism
group; this is the algebra of vector fields. Moreover, we must first
deal with the local properties of this algebra. Since any manifold is
locally diffeomorphic to $\Reals^N$, the only important parameter is
the dimensionality $N$, and it makes sense to talk about the algebra
of vector fields in $N$ dimensions, $Vect(N)$.  This algebra has
recently attracted some interest by physicists\ref{1-6}; references
some earlier mathematical literature can be found in Ref. 1.  The
one-dimensional case is special, because it admits a central extension
(Virasoro algebra), which is a cornerstone of modern theoretical
physics\ref{7}.  From the arguments above, it is clear that
$N$-dimensional local differential geometry may be considered as the
representation theory of $Vect(N)$, and it is in fact straightforward
to describe e.g. tensor fields and exterior derivatives in such terms.

In a recent paper\ref{1}, we discovered a new class of $Vect(N)$
representations which seem more natural than tensor fields, namely
conformal fields. A tensor field can be considered as a scalar field
decorated with indices from the rigid $gl(N)$ subalgebra. Similarly, a
conformal field has indices from the ``conformal'' subalgebra
$sl(N+1)$, which is obtained from $gl(N)$ by adding translations and
``conformal'' transformations. Because $gl(N) \subset sl(N+1)$, a
conformal field transforms nicely under a larger finite-dimensional
subalgebra than a tensor field does.

This paper is organized as follows. In section 2 we recollect some
relevant facts about tensor fields and exterior derivatives,
formulated in a fashion which emphasizes the representation theory
aspects. Section 3 contains the definition of conformal fields, in a
slightly more streamlined notation than in Ref. 1.  It also contains
some minor new results.  In section 4 we construct first-order
differential operators that are compatible with the $Vect(N)$ action.
These conformal exterior derivatives are considerably more abundant
than their tensor counterparts.  This leads to the development of a
form language, whose applications are investigated in section 5.
Because each conformal index can take one more value than the
corresponding tensor index, we propose in the final section that this
extra dimension can be interpreted as time.

\endchapter

\header{2. Tensor fields}

In some neighborhood of the origin, a vector field has the form
$f(x) \d^\mu$. By Fourier expansion we see that a basis of $Vect(N)$
is given
by $L^\mu(m) = \emx \d^\mu$, where $m = (m^1, \ldots, m^N)$ belongs
to an $N$-dimensional lattice $\Lambda$, and the brackets are
$$ [L^\mu(m), L^\nu(n)] = n^\mu L^\nu(m+n) - m^\nu L^\mu(m+n)
	\eqno(2.1) $$
Our convention is that the lattice is purely imaginary, which saves many
explicit references to the imaginary unit $i$.
Of special interest are $\Lambda = i\Ints^N$ and
$\Lambda = i\Reals^N$. The $\Lambda$-gradation expresses momentum
conservation.

An important class of $Vect(N)$ representations are tensor fields with
$p$ upper and $q$ lower indices and conformal weight $\lambda$, which are
constructed from $gl(N)$ representations as follows.
Assume that $\{T^\mu_\nu\}_{\mu,\nu=1}^N$ satisfies $gl(N)$, i.e.
$$ [T^\mu_\sigma, T^\nu_\tau] =
 \delta^\mu_\tau T^\nu_\sigma - \delta^\nu_\sigma T^\mu_\tau.
	\eqno(2.2) $$
Then it is easy to check that
$$L^\mu(m) = \emx \Big(\d^\mu + m^\sigma T^\mu_\sigma \Big)
	\eqno(2.3) $$
satisfies $Vect(N)$. This means for each $gl(N)$ representation we have a
corresponding $Vect(N)$ representation.

Instead of giving the matrix (2.3), we could equally well describe
the representation
by the action on a module. Let us elaborate on this trivial point, if
nothing else
because the signs have caused this author some confusion. Let $J^a$ be the
generators of some Lie algebra and $M^a$ a representation matrix,
$$ [J^a, J^b] = f^{ab}{}_c J^c, \qquad
[M^a, M^b] = f^{ab}{}_c M^c.
	\eqno(2.4) $$
Then the module is a vector space with basis $\phi$, and the Lie algebra
acts as
$J^a \phi = M^a \phi$ (representation indices suppressed).
Because $J^a$ only acts on $\phi$ and not on the numerical matrix $M^a$,
$$ [J^a, J^b] \phi = J^a M^b \phi - \axb = M^b M^a \phi - \axb
= -f^{ab}{}_c J^c \phi,
	\eqno(2.5) $$
so with this convention the structure constants change sign.

In particular, from the $gl(N)$ action on a tensor $\TT^p_q(\lambda)$,
$$ \eqalign {
T^\mu_\sigma \phi^\nup_\tauq = \lambda \delta^\mu_\sigma \phi^\nup_\tauq
+ \sum_{i=1}^p \delta^{\nu_i}_\sigma \phi^{\argsx\nu p \mu}_\tauq
- \sum_{j=1}^q \delta^\mu_{\tau_j} \phi^\nup_{\argsx\tau q\sigma},
}	\eqno(2.6) $$
we find that the corresponding action of $Vect(N)$ on the tensor field
$\phi^\nup_\tauq(x) = \phi^\nup_\tauq \otimes f(x)$ is
$$ \eqalign {
L^\mu(m) \phi^\nup_\tauq(x) =
\emx \bigg( &(\d^\mu + \lambda m^\mu) \phi^\nup_\tauq(x) \cr
&+ \sum_{i=1}^p m^{\nu_i} \phi^{\argsx\nu p \mu}_\tauq(x)
- \sum_{j=1}^q \delta^\mu_{\tau_j} m^\sigma
\phi^\nup_{\argsx\tau q\sigma} \bigg).
}	\eqno(2.7) $$
By abuse of notation, we also denote this $Vect(N)$ module by
$\TT^p_q(\lambda)$.
The parameter $\lambda$ will be refered to as the
{\it (conformal) weight}. Note that
in one dimension the relevent quantity is $\lambda + p - q$.

A slight generalization of tensor fields is obtained by shifting
$\d^\mu \to \d^\mu + h^\mu$. The constant vector $h^\mu$ is of course
only defined
modulo $\Lambda$, because otherwise it could be absorbed into a relabelling of
the Fourier components. E.g.,
the action on a scalar field now reads
$$L^\mu(m) \phi = \emx (\d^\mu + \lambda m^\mu + h^\mu) \phi.
	\eqno(2.8) $$
By forgetting the vector index we recognize this as the transformation
law of a
primary field in conformal field theory. In Ref. 1 it was suggested that
$L^\mu(0)$ generates rigid translations in space-time, and thus $h^N$
could be interpreted as a mass. In view of the current section 6, this
assumption
seems incorrect. Moreover, for continuous $\Lambda$, $h^\mu$ can be eliminated
altogether by substituting $\phi \to \exp(-h \ddot x) \phi$.

Recall that a {\it homomorphism}, or {\it module map}, is a map
between modules that preserves the Lie algebra structure. If the
generic Lie algebra (2.4) has two matrix representations $M^a$ and
$N^a$, $d$ is a homomorphism if
$$ J^a (d\phi) = d J^a \phi = d M^a \phi = N^a d \phi,
	\eqno(2.9) $$
i.e. $d M^a = N^a d$.

Two important classes of homomorphisms can be defined for tensor fields.
The first is pointwise multiplication, which is the map
$$\eqalign{
\TT^{p_\phi}_{q_\phi}(\lambda_\phi) \times
\TT^{p_\psi}_{q_\psi}(\lambda_\psi)
&\longrightarrow
\TT^{p_\phi+p_\psi}_{q_\phi+q_\psi}(\lambda_\phi+\lambda_\psi) \cr
\phi(x) \times \psi(y) &\mapsto (\phi\psi)(x) = \phi(x) \psi(x)
}	\eqno(2.10) $$
Note that this module is a much smaller than the tensor product
$\phi(x) \psi(y)$.
The result (2.10) follows from Leibniz' rule,
$$\d^\mu = \d^\mu \otimes 1 + 1 \otimes \d^\mu,	\qquad
T^\mu_\nu = T^\mu_\nu \otimes 1 + 1 \otimes T^\mu_\nu.
	\eqno(2.11) $$
In the case that $\phi$ and $\psi$ are two scalar fields with non-zero
conformal weights,
$$ \eqalign{
L^\mu(m) (\phi \psi) &= (\emx (\d^\mu + \lambda_\phi m^\mu) \phi) \psi
+ \phi (\emx (\d^\mu + \lambda_\psi m^\mu) \psi) \cr
&= \emx (\d^\mu + (\lambda_\phi + \lambda_\psi) m^\mu) (\phi \psi),
}	\eqno(2.12) $$
which proves that $\phi\psi$ is a new scalar field with weight
$\lambda_\phi + \lambda_\psi$.

The second homomorphism is the {exterior derivative}. Tensor fields
with several indices admit submodules which are symmetric or skew in
some indices. Of particular interest is the module $\Omega^p \subset
\TT^p_0(0)$ consisting of totally skew $p$-tensors ({\it forms}),
because then we can define the sequence of maps $d_p$,
$$ \Omega^0 \mapright{d_0} \Omega^1 \mapright{d_1} \ldots
\Omega^N \mapright{d_N} 0,
	\eqno(2.13) $$
satisfying $d_{p+1} d_p = 0$.
Explicitly, the maps are given by
$$ (d_p \phi)^{\nu_1\ldots\nu_{p+1}} =
\pefac \d^{[\nu_1} \phi^{\nu_2\ldots\nu_{p+1}]},
	\eqno(2.14) $$
where the bracketed indices are skew-symmetrized. To illustrate how to
prove that $d_p$ are maps, and also to see what goes wrong if we apply
it to tensor fields which are not forms, consider applying $d_0$ to a
scalar field with conformal weight $\lambda$. The action of $Vect(N)$ is
$$ \eqalign{
L^\mu (d\phi)^\nu &= \d^\nu \emx (\d^\mu + \lambda m^\mu) \phi \cr
&= \emx \bigg( (\d^\mu + \lambda m^\mu) \d^\nu \phi + m^\nu \d^\mu \phi
+ \lambda m^\mu m^\nu \phi \bigg)
}	\eqno(2.15) $$
Thus, $(d\phi)^\nu$ transforms as a tensor apart from the last term.
However, this
term is absent if $\lambda = 0$, which is precisely the condition that
$\phi \in \Omega^0$. The verification of the higher maps is straightforward.

We also have a dual sequence involving {\it chains}
$\Omega_p \subset \TT^0_p(1)$.
$$ \Omega_N \mapright{\bar d_N} \Omega_{N-1}
\mapright{\bar d_{N-1}} \ldots
\mapright{\bar d_1} \Omega_0.
	\eqno(2.16) $$
These maps are given by
$$ (\bar d_p \bar \phi)_{\nu_1\ldots\nu_{p-1}}
= \d^\sigma \bar \phi_{\nu_1\ldots\nu_{p-1}\sigma}
	\eqno(2.17) $$
and again $\bar d_{p-1} \bar d_p = 0$. Forms and chains are dual
in the sense that an invariant pairing can be defined by
$$ \langle \phi(x), \bar\phi(y) \rangle = \delta(x-y)
	\eqno(2.18) $$
and $d_p$ and $\bar d_p$ are dual maps relative to this pairing.

Finally, pointwise multiplication endows forms with a ring structure
which is compatible with the exterior derivative. The multiplication
is given by the exterior product
$\wedge: \Omega^{p} \times \Omega^{q} \to \Omega^{p+q}$
$$ (\phi \wedge \psi)^{\nu_1\ldots\nu_p\nu_{p+1}\ldots\nu_{p+q}}
= \pqfac \phi^{[\nu_1\ldots\nu_p} \psi^{\nu_{p+1}\ldots\nu_{p+q}]}.
	\eqno(2.19) $$
The compatibility is expressed by Leibniz' rule,
$$ d_{p+q} (\phi \wedge \psi)
= (d_p \phi) \wedge \psi + (-)^p \phi \wedge (d_q \psi).
	\eqno(2.20) $$
No similar map can be constructed for chains.

To distinguish objects with tensor indices from those derived from the
conformal fields introduced below, we sometimes prefix them with the
word ``tensor'' and talk about tensor forms, tensor vielbeins, etc.

\endchapter

\header{3. Conformal fields}

The main result of Ref. 1 was the discovery of a new class of
$Vect(N)$ representations called {\it conformal fields}. If $T^A_B$,
$A, B = 0, 1, \ldots N$ are the generators of $gl(N+1)$, i.e.
$$ [T^A_B, T^C_D] = \delta^A_D T^B_C - \delta^B_C T^A_D,
	\eqno(3.1) $$
then
$$ L^\mu(m) = \emx \bigg( \d^\mu + (m^B + k^B) T^\mu_B
+ c m^\mu m^B T^A_B x_A \bigg).
	\eqno(3.2) $$
satisfies $Vect(N)$. Here
$$ \eqalign{
m^A \equiv (m^0, m^\mu) &= (- m \ddot x, m^\mu), \cr
x_B \equiv (x_0, x_\nu) &= (1, x_\nu),	\cr
k^A \equiv (k^0, k^\mu) &= (1, 0),
}	\eqno(3.3) $$
and $c$ is a c-number parameter.

By introducing the $N+1$-dimensional derivative
$\d^A = (- x \ddot \d, \d^\mu)$, the
relevant algebraic properties can be summarized as
$$ x_A m^A = x_A \d^A  = 0, \qquad
k^A x_A = 1, \qquad
\d^A x_A = N .
	\eqno(3.4) $$
$$ \eqalign{
[\d^A, \d^B] &= k^A \d^B - k^B \d^A \cr
[\d^A, x_B] &= \delta^A_B - k^A x_B
} \qquad \eqalign{
[\d^A, m^B] &= - k^B m^A \cr
[\d^A, \emx] &= m^A \emx
}	\eqno(3.5) $$
and all commutators between $m^A$, $x_B$ and $k^C$ vanish.

By substituting different representations of $gl(N+1)$ into (3.2), we
obtain various representations of $Vect(N)$. In particular, if
$\{\phi^\Ap_\Bq\}_{A_i,B_j=1}^{N+1}$ is a basis for the $gl(N+1)$
module $\TT^p_q(\lambda)$ (2.6), the action of
$Vect(N)$ on $\phi^\Ap_\Bq(x) = \phi^\Ap_\Bq \otimes f(x)$ is given by
$$ \eqalign {
L^\mu(m) \phi^\Ap_\Bq(x) &=
\emx \bigg( (\d^\mu + \lambda m^\mu) \phi^\Ap_\Bq(x) \cr
&+ \sum_{i=1}^p \Big\{ (m^{A_i} + k^{A_i}) \phi^{\argsx A p \mu}_\Bq(x)
+ c m^\mu m^{A_i} x_C \phi^{\argsx A p C}_\Bq(x) \Big\} \cr
&- \sum_{j=1}^q \Big\{ \delta^\mu_{B_j} (m^C + k^C)
\phi^\Ap_{\argsx B q C}(x)
+ c m^\mu m^C x_{B_j} \phi^\Ap_{\argsx B q C}(x) \Big\}  \bigg)
}	\eqno(3.6) $$
This defines the {\it conformal field} $\CC^p_q(\lambda, c)$.

The discovery (and name) of conformal fields arose in Ref. 1 from a
study of of the largest finite-dimensional subalgebra of $Vect(N)$,
which is the ``conformal'' algebra $sl(N+1)$. For a scalar field,
$$ J^A_B \equiv \pmatrix{J^0_0 & J^0_\nu \cr J^\mu_0 & J^\mu_\nu}
= \pmatrix{-x\ddot\d & -x_\nu x\ddot\d \cr \d^\mu & x_\nu \d^\mu}
= x_B \d^A
	\eqno(3.7) $$
Generally, the conformal generators are found by differentiating $L^\mu(m)$
with respect to $m$.
$$ J^A_B = \pmatrix{
\displaystyle{-{{\d L^\sigma(m)} \over {\d m^\sigma}} \bigg|_{m=0}}
&\displaystyle{-{{\d^2 L^\sigma(m)} \over {\d m^\nu \d m^\sigma}}
\bigg|_{m=0}} \cr\cr
{L^\mu(0)}
& \displaystyle{{{\d L^\mu(m)} \over {\d m^\nu}} \bigg|_{m=0}}
}.
	\eqno(3.8) $$
$sl(N+1)$ is not conformal in the usual sense that it preserves
angles, because in order to define an angle a metric has to be
introduced in an {\it ad hoc} fashion. Rather, it is a ``conformal''
algebra in the profane sense that the generators are of the form $\d$,
$x\d$ and $x^2\d$.

The procedure to arrive at (3.8) may seem dubious, but it is merely a
consequence of our particular basis. $Vect(N)$ can be formulated in
terms of generators $L(f_\mu)$, $f_\mu(x)$ some vector-valued function.
$$ [L(f_\mu), L(g_\mu)] = L(f \ddot \d g_\mu) - L(g \ddot \d f_\mu)
	\eqno(3.9) $$
With such a basis, the conformal subalgebra is completely natural.

Conformal fields with $c=-1/(N+1)$ are singled out when the restriction to
$sl(N+1)$ is considered. Using the formula
$$ {{\d m^A} \over {\d m^\nu}}
= (-x_\nu, \delta^\mu_\nu)
= \delta^A_\nu - k^A x_\nu
	\eqno(3.10) $$
we find
$$ J^A_B = x_B \d^A + T^A_B - k^A x_B T^C_C
	\eqno(3.11) $$
and in particular $J^A_A = 0$, as expected. For other values of $c$
we still of course
obtain a representation of $sl(N+1)$, but it is more messy.

Just as tensor fields, conformal fields admit a pointwise tensor product.
This is
because Leibniz' rule holds both for the derivative $\d^\mu$ and for the
$gl(N+1)$ generator $T^A_B$. Thus, we have the map
$$ \eqalign{
\CC^{p_1}_{q_1}(\lambda_1, c) \times \CC^{p_2}_{q_2}(\lambda_2, c)
&\longrightarrow \CC^{p_1+p_2}_{q_1+q_2}(\lambda_1+\lambda_2, c) \cr
\phi(x) \times \psi(y) &\mapsto (\phi\psi)(x) = \phi(x) \psi(x).
}	\eqno(3.12) $$
A point worth noting is that the parameter $c$ must be the same in
both participating fields. In this sense (3.12) is similar to a fusion
rule in conformal field theory, and $c$ takes the role of the central
charge. We believe that this similarity is not coincidental.

{}From (3.2) it immediately follows that
$\CC^0_0(\lambda, c) \equiv \TT^0_0(\lambda)$.
More generally, we may ask if it is possible to construct a conformal
field from a tensor field. This construction is carried out for vector
fields only, but it can be applied to arbitrary tensors by treating
each index separately.
$$\eqalign{
\phi^\mu \in \TT^1_0(0) &\implies
\phi^A = (-x_\sigma, \delta^\mu_\sigma) \phi^\sigma \equiv
 (-x \ddot \phi, \phi^\mu) \in \CC^1_0(0,c) \cr
\psi \in \TT^0_0(0) &\implies
\psi_B = x_B \psi \equiv (\psi, x_\nu \psi) \in \CC^0_1(0,c) \cr
}	\eqno(3.13) $$
We say that such conformal fields are {\it tensor derived}. However,
not every conformal field is tensor derived, because the definition
(3.13) implies certain relations, namely
$$ \eqalign{
x_A \phi^A &= -x_0 x \ddot \phi + x_\mu \phi^\mu = 0, \cr
m^B \psi_B & = 0, \cr
k^B \psi_B &\in \TT^0_0(0)
}	\eqno(3.14) $$
The parameter $c$ is irrelevant for tensor derived fields,
because (3.14) guarantees that the term multiplying $c$ vanishes.
Let us prove that $\psi_B$ does transform as claimed.
$$ \eqalign{
L^\mu(m) \psi_B &= x_B \emx \d^\mu \psi \cr
&= \emx \big( \d^\mu(x_B \psi) - \delta^\mu_B \psi \big) \cr
&= \emx \big( \d^\mu \psi_B - \delta^\mu_B (m^C + k^C)\psi_C
 + c m^\mu x_B m^C \psi_C \big)
}	\eqno(3.15) $$
This shows that tensor fields are intimately related to a particular
class of conformal fields. Henceforth, we focus on conformal fields
that are not tensor derived.

\endchapter

\header{4. Conformal forms and exterior derivatives}

In this section we construct the conformal analog of forms.

The $Vect(N)$ module $\Omega^p(\lambda, c) \subset \CC^p_0(\lambda, c)$
consists of conformal fields which are skew in all $p$ upper
indices. Similarly, the elements in the module $\Omega_p(\lambda, c)
\subset \CC^0_p(\lambda, c)$ are skew in all $p$ lower indices. These
objects will be refered to as positive and negative {\it conformal
forms}, respectively.  The form degree will often be indicated by the
notation $\phi_p$ and $\psin_{-q}$, as shorthand for $\phi^\Ap(x)$ and
$\psin_\Ap(x)$, respectively.  As in section 2, bracketed indices are
anti-symmetrized. Our convention is that $\phi^{[\Ap]} = p! \phi^\Ap$
for a field that is already skew.

\proclaim{Theorem I}
{\sl
There is a map ({\it conformal exterior derivative})
$$ \eqalign{
d_p(\lambda,c): \quad
\Omega^p(\lambda, c) &\longrightarrow \Omega^{p+1}(\lambda, c) \cr
(\phi_p)^{A_1 \ldots A_p} &\mapsto
(d_p(\lambda,c)\phi_p)^{A_1 \ldots A_{p+1}}
\equiv {1 \over {(p+1)!}} (\d^{[A_1} + \gamma_p(\lambda,c) k^{[A_1})
(\phi_p)^{A_2 \ldots A_{p+1}]}
}	\eqno(4.1) $$
where $\gamma_p(\lambda,c) = \lambda/c - p$.

There is a map
$$ \eqalign{
\dn_p(\lambda,c): \quad
\Omega_p(\lambda, c) &\longrightarrow \Omega_{p-1}(\lambda, c) \cr
(\phin_{-p})_{A_1 \ldots A_p}
&\mapsto (\dn_p(\lambda,c)\phin_{-p})_{A_1 \ldots A_{p-1}}
\equiv (\d^B + \gamman_p(\lambda,c) k^B) (\phin_{-p})_{A_1 \ldots A_{p-1}B}
}	\eqno(4.2) $$
where $\gamman_p(\lambda,c) = (\lambda-1)/c + p - N - 1$.

These maps satisfy $d_{p+1} d_p = 0$ and $\dn_{p-1} \dn_p = 0$.
Although $\Omega_0(\lambda,c) \equiv \Omega^0(\lambda,c)$, $d_0 \dn_1 \ne 0$.
}
\endproclaim

For $c = -1/(N+1)$, $\gamman_p(\lambda,c) = \gamma_{-p}(\lambda,c)$.
Recall that this was the value of $c$ which simplified the
inherited $sl(N+1)$ representation.
The normalization is fixed by compatibility with the wedge product below.

The following diagram summarizes the situation.
$$ \eqalign{
&\matrix{
&\Omega^0(\lambda,c) &\mapright{d_0}
&\Omega^1(\lambda,c) &\mapright{d_1} &\ldots
&\Omega^{N-1}(\lambda,c) &\mapright{d_{N-1}}
&\Omega^N(\lambda,c) &\mapright{d_N} &0
\cr & || \cr
&\Omega_0(\lambda,c) &\mapleft{\dn_1}
&\Omega_1(\lambda,c) &\mapleft{\dn_2} &\ldots
&\Omega_{N-1}(\lambda,c) &\mapleft{\dn_N}
&\Omega_N(\lambda,c)
}
}	\eqno(4.3) $$

\demo{Proof}
Let $\phi \in \Omega^0(\lambda, c) \equiv \CC^0_0(\lambda,c)$.
Then $(d\phi)^A$ transforms as
$$ \eqalign{
L^\mu(m) (d\phi)^A
&= (\d^A + \gamma k^A) \emx (\d^\mu + \lambda m^\mu) \phi \cr
&= \emx \bigg( m^A (\d^\mu + \lambda m^\mu) \phi + k^A \d^\mu \phi
+ (\d^\mu + \lambda m^\mu) (\d^A + \gamma k^A) \phi \bigg) \cr
&= \emx \bigg(  (\d^\mu + \lambda m^\mu) (d\phi)^A
+ (m^A + k^A) (d\phi)^\mu
+ \lambda m^\mu {1 \over \gamma} x_B (d\phi)^B \bigg)
}	\eqno(4.4) $$
In the last line, we used that
$(d\phi)^\mu = \d^\mu \phi$ ($k^\mu = 0$) and
$x_B (d\phi)^B = \gamma \phi$. Eq. (4.4) is the transformation law of
$\Omega^1(\lambda, c)$ provided that we identify $c = \lambda/\gamma$.

If $\phi^A \in \Omega^1(0, c) \equiv \CC^1_0(1,c)$, $(d\phi)^{AB}$
transforms as
$$ \eqalign{
&L^\mu(m) (d\phi)^{AB} \cr
&= \half (\d^A + \gamma k^A)
\emx \bigg( (\d^\mu + \lambda m^\mu) \phi^B + (m^B + k^B) \phi^\mu
+ c m^\mu m^B x_C \phi^C \bigg) - \AxB \cr
&= \half \emx \bigg( m^A \big( (\d^\mu + \lambda m^\mu) \phi^B
+ (m^B + k^B) \phi^\mu
+ c m^\mu m^B x_C \phi^C \big) + k^A \d^\mu \phi^B
- k^B m^A \phi^\mu \cr
&\qquad+ c m^\mu( -k^B m^A x_C \phi^C + m^B \phi^A - m^B k^A x_C \phi^C)
+ (\d^\mu + \lambda m^\mu) (\d^A + \gamma k^A) \phi^B \cr
&\qquad+ (m^B + k^B) (\d^A + \gamma k^A) \phi^\mu
+ c m^\mu m^B x_C (\d^A + \gamma k^A) \phi^C \bigg) - \AxB \cr
&= \emx \bigg(  (\d^\mu + \lambda m^\mu) (d\phi)^{AB}
+ (m^A + k^A) (d\phi)^{\mu B} + (m^B + k^B) (d\phi)^{A\mu} \cr
&\qquad + {c\over2} m^\mu m^B \big(x_C (\d^A + \gamma k^A) \phi^C +
(1 - {\lambda \over c}) \phi^A - \AxB \big) \bigg).
}	\eqno(4.5) $$
If we now realize that
$$ 2 x_C (d\phi)^{AC} =  x_C (\d^A + \gamma k^A) \phi^C - \gamma \phi^A,
	\eqno(4.6) $$
the last term becomes
$ c m^\mu m^B x_C (d\phi)^{AC} - \AxB$
provided that $\gamma + 1 - \lambda/c = 0$, which precisely is the
condition for $\gamma_1$.

Let $\phi_A \in \Omega_1(\lambda, c) \equiv \CC^0_1(\lambda,c)$.
Then $\dn\phi$ transforms as
$$ \eqalign{
L^\mu(m) (\dn\phi)
&= (\d^A + \gamman k^A) \emx \bigg((\d^\mu + \lambda m^\mu) \phi_A
- \delta^\mu_A (m^B + k^B)\phi_B - c m^\mu x_A m^B \phi_B \bigg) \cr
&= \emx \bigg( m^A \big((\d^\mu + \lambda m^\mu) \phi_A
- \delta^\mu_A (m^B + k^B)\phi_B - c m^\mu x_A m^B \phi_B \big)
+ k^A \d^\mu \phi_A \cr
&\qquad - \delta^\mu_A (-m^A) k^B \phi_B
- c m^\mu (N m^B \phi_B - x_A m^A k^B \phi_B)
+ (\d^\mu + \lambda m^\mu) \dn\phi \cr
&\qquad - \delta^\mu_A (m+k)^B (\d^A + \gamman k^A) \phi_B
- c m^\mu x_A m^B (\d^A + \gamman k^A) \phi_B \bigg) \cr
&= \emx \bigg(  (\d^\mu + \lambda m^\mu) \dn\phi
+ m^\mu (\lambda - 1 - Nc - \gamman c) m^A \phi_A \bigg).
}	\eqno(4.7) $$
This is the transformation law of $\Omega_0(\lambda, c)$ provided
that the last term
vanishes, which yields the condition on $\gamman$.

The same brute force method can be applied to higher conformal forms.
When this is done it is realized that the structure is such that the
higher exterior derivatives are module maps. However, a simpler proof
can be constructed, at least for positive forms, but it must be
postponed until Leibniz' rule has been verified.

Two consequtive maps give zero. If $\phi \in \Omega^p$,
$$ (d_{p+1} d_p \phi)^{ABC_1\ldots C_p}
\propto (\d^{[A} + \gamma_{p+1} k^{[A}) (\d^{B} + \gamma_p k^{B})
\phi^{C_1\ldots C_p]}
	\eqno(4.8) $$
Most terms vanish because of anti-symmetry in $A$ and $B$, but two remain:
$$ \d^{[A} \d^{B]} = [\d^A, \d^B] = k^A \d^B - k^B \d^A
	\eqno(4.9) $$
and
$$ \gamma_{p+1} k^{[A} \d^{B]} + \gamma_p \d^{[A} k^{B]}
 = (\gamma_{p+1} - \gamma_p) k^{[A} \d^{B]}
	\eqno(4.10) $$
However, because $\gamma_{p+1} - \gamma_p = -1$, these terms cancel and
$d_{p+1} d_p \phi = 0$.

Similarly, if  $\phi \in \Omega_{p+2}$,
$$ (\dn_{p+1} \dn_{p+2} \phi)_{C_1\ldots C_p}
= (\d^A + \gamman_{p+1} k^A) (\d^B + \gamman_{p+2} k^B)
\phi_{C_1\ldots C_p AB},
	\eqno(4.11) $$
which is identically zero because $\gamman_{p+2} - \gamman_{p+1} = 1$.
Remains to
prove that $d_0 \dn_1 \ne 0$. But this is obvious, because in the expression
$$ (d_0 \dn_1 \phi)^A = (\d^A + \gamma_0 k^A) (\d^B + \gamman_1 k^B) \phi_B
	\eqno(4.12) $$
nothing cancels.
\qed
\enddemo

Conformal forms can be endowed with a ring
structure by introducing the following products.
$$ \eqalign{
\wedge: \qquad &\Omega^p(\lambda_\phi, c) \times  \Omega^q(\lambda_\psi, c)
\longrightarrow  \Omega^{p+q}(\lambda_\phi+\lambda_\psi, c) \cr
&(\phi_p \wedge \psi_q)^{\Ap \Bq}(x)
= {1 \over {(p+q)!}} \phi^{[\Ap}(x) \psi^{\Bq]}(x) \cr
\: : \qquad &\Omega^p(\lambda_\phi, c) \times  \Omega_q(\lambda_\psi, c)
\longrightarrow
\Omega_{q-p}(\lambda_\phi+\lambda_\psi, c) \qquad (p \le q) \cr
&(\phi_p \: \psin_{-q})_{A_1 \ldots A_{q-p}} (x)
= \phi^\Bp(x) \psin_{A_1 \ldots A_{q-p} \Bp}(x)
}	\eqno(4.13) $$

The wedge product $\wedge$ is clearly associative, and the
contraction product $\:$
associates with $\wedge$ in the following sense.
$$ (\phi_p \wedge \psi_q) \: \thetan_{-r}
= \phi_p \: (\psi_q \: \thetan_{-r})
= (-)^{pq} \psi_q \: (\phi_p \: \thetan_{-r})
	\eqno(4.14) $$
where $p+q \le r$. Conformal forms commute as usual.
$$ \psi_q \wedge \phi_p = (-)^{pq} \phi_p \wedge \psi_q
	\eqno(4.15) $$
Note that our notation differs from Ref. 1, where the double dots were used
to indicate contraction of arbitrary conformal indices. Here its use
is restricted to forms.

The contraction product could naturally be extended to the case that
combined form degree is positive. A simple example would be
$$ (\phi_2 \: \psin_{-1})^A = \phi^{AB} \psin_B
	\eqno(4.16) $$
However, it is easy to see that this map would violate associativity.
$$ (\phi_p \wedge \psi_q) \: \thetan_{-r}
\ne \phi_p \: (\psi_q \: \thetan_{-r})
	\eqno(4.17) $$
if $p+q > r$. Moreover, it turns out that this extension would not comply with
Leibniz' rule, e.g.
$$ d_0(\phi_1 \: \psin_{-1}) \ne
(d_1 \phi_1) \: \psin_{-1} - \phi_1 \: (\dn_1 \psin_{-1})
	\eqno(4.18) $$

To obtain a higher degree of symmetry between positive and negative
forms, it would also be tempting to introduce a multiplication rule
for negative forms. However, we have failed to do so. The obvious
candidate,
$$\eqalign{
\vee: \qquad &\Omega_p(\lambda_\phi, c) \times  \Omega_q(\lambda_\psi, c)
\longrightarrow  \Omega_{p+q}(\lambda_\phi+\lambda_\psi, c) \cr
&(\phin_{-p} \vee \psin_{-q})_{A_1 \ldots A_p B_1 \ldots B_q}(x)
= \phin_{[A_1 \ldots A_p}(x) \psin_{B_1 \ldots B_q]}(x) \cr
}	\eqno(4.19) $$
is certainly a module map, but Leibniz' rule does not hold for the
exterior derivative and this product.

\proclaim{Theorem II} (Leibniz' rule)
{\sl
The exterior derivative is compatible with the wedge and contraction products
in the following sense.
$$ \eqalign{
d_{p+q}(\lambda_\phi+\lambda_\psi, c) (\phi_p \wedge \psi_q)
&= d_p(\lambda_\phi, c)\phi_p \wedge \psi_q
+ (-)^p \phi_p \wedge d_q(\lambda_\psi, c) \psi_q \cr
\dn_{q-p}(\lambda_\phi+\lambda_\psi, c) (\phi_p \: \psin_{-q})
&= d_p(\lambda_\phi, c)\phi_p \: \psin_{-q}
+ (-)^p \phi_p \: \dn_q(\lambda_\psi, c) \psin_{-q}
}	\eqno(4.20) $$
}
\endproclaim

\demo{Proof}
We want to prove that
$$ \eqalign{
&\pqefac (\d^{[C} + \gamma_{p+q}(\lambda_\phi+\lambda_\psi, c) k^{[C})
\pqfac (\phi^{[\Ap}  \psi^{\Bq]]}) \cr
= &\pqefac (\d^{[C} + \gamma_{p+q}(\lambda_\phi+\lambda_\psi, c) k^{[C})
 (\phi^{\Ap}  \psi^{\Bq]})
}	\eqno(4.21) $$
equals
$$ \eqalign{
& \pqefac \big(\pefac (\d^{[[C} + \gamma_p(\lambda_\phi, c) k^{[[C})
 \phi^{\Ap]} \big) \psi^{\Bq]} \cr
&+ (-)^p \pqefac \phi^{[\Ap}
\big(\qefac (\d^{[C} + \gamma_q(\lambda_\psi, c) k^{[C}) \psi^{\Bq]]} \big)
}	\eqno(4.22) $$
But this is true because Leibniz' rule holds for the partial derivative,
the factors
$1/(p+1)!$ and $1/(q+1)!$ cancel the extra anti-symmetrization, and
$$\gamma_{p+q}(\lambda_\phi+\lambda_\psi, c)
= \gamma_p(\lambda_\phi, c) + \gamma_q(\lambda_\psi, c).
	\eqno(4.23) $$

To prove the second part, we note that
$$
(d(\phi_p \: \psin_{-q}))_\Aqp
= (\d^C + \gamman_{q-p}(\lambda_\phi+\lambda_\psi, c) k^C)
(\phi^\Bp \psin_{\Aqp C \Bp})
	\eqno(4.24) $$
whereas
$$ \eqalign{
&(d\phi_p \: \psin_{-q})_\Aqp + (-)^p (\phi_p \: d\psin_{-q})_\Aqp \cr
= &\pefac ((\d^{[C} + \gamma_p(\lambda_\phi, c) k^{[C})
\phi^{\Bp]}) \psin_{\Aqp C \Bp} \cr
& \qquad + (-)^p \phi^\Bp (\d^C + \gamman_q(\lambda_\psi, c) k^C)
\psin_{\Aqp \Bp C})
}	\eqno(4.25) $$
We now use that $\psin_{\Aqp \Bp C} = (-)^p \psin_{\Aqp C \Bp}$ and
that the first term consists of $(p+1)!$ identical terms, which
cancels the factor in the denominator. It then follows that (4.24) and
(4.25) agree provided that
$$ \gamman_{q-p}(\lambda_\phi+\lambda_\psi, c)
= \gamma_p(\lambda_\phi, c) + \gamman_q(\lambda_\psi, c).
	\eqno(4.26) $$
{}From the definition of $\gamma$ and $\gamman$ in Theorem I it is clear
that this equation holds identically.
\qed
\enddemo

As promised in the proof of Theorem I,
Leibniz' rule can be utilized to establish the existence of the exterior
derivative for higher forms.
Let $\phi_p \in \Omega^p(\lambda,c)$ and $\psi_q \in \Omega^q(0,c)$,
which means that
$\phi_p \wedge \psi_q \in \Omega^{p+q}(\lambda,c)$.
Assume that we have established that
$d_p \phi_p$ and $d_q \psi_q$ transform as conformal fields. Then
$$ d_{p+q} (\phi_p \wedge \psi_q)
= (d_p \phi_p) \wedge \psi_q + (-)^p \phi_p \wedge (d_q \psi_q)
	\eqno(4.27) $$
must transform as a conformal field of the appropriate type because
both terms to the right do so. This argument extends to linear
combinations of the same type.  Since any form is a linear combination
of exterior products and the base case is clear from the explicit
calculation (4.5), we conclude by induction that higher positive forms
do exist.

The existence of negative forms can be motivated, although not
strictly proved, by using Leibniz' rule backwards. The relation
$$ \dn_{q-p} (\phi_p \: \psin_{-q})
= (d_p \phi_p) \: \psin_{-q} + (-)^p \phi_p \: (\dn_q \psin_{-q})
	\eqno(4.28) $$
can be viewed as an equation for $\dn_q \psin_{-q}$. Because all other
entities in (4.28) are conformal fields, this is a strong hint that
the solution is conformal; that it is skew is manifest. We have
explicitly verified that $\dn_2$ is a module map.

\endchapter

\header{5. Covariant derivatives and conformal vielbeins}

Having established a form language and the existence of exterior
derivatives, we can now apply the standard machinery of local
differential geometry\ref{8,9} to conformal forms.  Since the
algebraic manipulations leading to the formulas in this section are
standard, we do not spell out each of them in detail.  However, it
should be stressed that the results themselves are new, because the
formalism is applied to conformal forms rather than to tensor forms.

Consider the algebra
of maps from $N$-dimensional space to a finite-dimensional Lie algebra $\oj$,
$Map(N, \oj)$ (algebra of gauge transformations).
$$ [J^a(m), J^b(n)] = f^{abc} J^c(m+n),
	\eqno(5.1) $$
where $f^{abc}$ are the totally skew-symmetric structure constants of
$\oj$ and $m, n \in \Lambda$. Moreover, we only consider Lie algebras
with a non-denerate Killing metric $\delta^{ab}$, so there is no need
to distinguish between upper and lower $\oj$ indices.  The gauge
algebra must be compatible with $Vect(N)$ in the sense that
$$ [L^\mu(m), J^b(n)] = n^\mu J^c(m+n)
	\eqno(5.2) $$

An explicit representation of (5.2) is given by
$$ J^a(m) = \emx M^a,
	\eqno(5.3) $$
where $M^a$ are matrices in a finite-dimensional $\oj$ irrep.
In other words, we have a $Map(N, \oj)$ module consisting of
a $\oj$-representation-valued conformal field, and action
$$ J^a(m) \phi(x) = \emx M^a \phi(x)
	\eqno(5.4) $$
The representation (5.4) is compatible with $Vect(N)$.

As usual, the exterior derivative does not commute with the action of
the gauge algebra, so it has to be covariantized.  A {\it gauge
connection} is a $\CC^1_0(0,c)$ conformal field transforming under
$\oj$ as
$$ [J^a(m), \omega^{Bb}(x)]
= \emx \Big( - f^{abc} \omega^{Bc}(x) - m^B \delta^{ab} \Big)
	\eqno(5.5) $$
Thus it transforms according to the adjoint representation apart from an
inhomogeneous term. The gauge connection can be viewed as a conformal 1-form
$\omega_1^b \in \Omega^1(0,c)$. In form language, (5.5) takes the form
$$ [J^a(m), \omega_1^b]
= \emx \Big( - f^{abc} \omega_1^b - \delta^{ab} m_1 \Big),
	\eqno(5.6) $$
where the constant $m^A$ is regarded as a 1-form $m_1$.

The {\it covariant derivative} of a form transforming as (5.3) is
$$ \eqalign{
D_p(\omega,\lambda,c) \phi_p(x)
&= d_p(\lambda,c) \phi_p(x) + M^b \omega_1^b \wedge \phi_p(x). \cr
\Dn_p(\omega,\lambda,c) \psin_{-p}(x)
&= \dn_p(\lambda,c) \psin_{-p}(x) + M^b \omega_1^b \: \psin_{-p}(x).
}	\eqno(5.7) $$
E.g., for 0-forms we have explicitly
$$ (D_0\phi)^A = (\d^A + {\lambda \over c} k^A) \phi + M^b \omega^{Ab} \phi
	\eqno(5.8) $$
That this is a $Vect(N)$ map is clear because both terms to the right are
separately so. Remains to check the gauge part.
$$ \eqalign{
J^a(m) D_p \phi_p
&= (d_p + \omega_1^b M^b) \emx M^a \phi_p +
\emx ( -f^{abc} \omega_1^c - \delta^{ab} m_1) \wedge M^b \phi_p \cr
&= \emx ( M^a D_p \phi_p - M^a m_1 \wedge \phi_p)
+  M^a [d_p, \emx] \wedge \phi_p
}	\eqno(5.9) $$
Noting that $ [d_p, \emx] = \emx m_1$ for every exterior derivative
contructed in the previous section, we obtain the claimed result.

The combination of two covariant derivatives gives
$$ D_{p+1}(\omega,\lambda,c) \, D_p(\omega,\lambda,c) \, \phi_p
= M^a R_2^a(\omega,c) \wedge \phi_p,
	\eqno(5.10) $$
where the {\it field strength} is the $\Omega_2(0,c)$ form
$$ R_2^a(\omega,c) = D_1(\omega,0,c) \omega_1^a
= d_1(0,c) \omega_1^a - \half f^{abc} \omega_1^b \wedge \omega_1^c.
	\eqno(5.11) $$
It is subject to the Bianchi identity
$$ D_2(\omega,0,c) R_2^a(\omega,c) \equiv 0.
	\eqno(5.12) $$

A {\it frame algebra} is a special type of gauge algebra that admits a
{\it conformal vielbein}, which is a $\CC^1_0(0,c)$ field $e^{Ai}$
with a frame vector index. The vielbein is defined by the non-trivial
property of having a two-sided inverse $\en^i_A$ everywhere.
$$ e^{Ai}(x) \en^i_B(x) = \delta^A_B, \qquad
e^{Ai}(x) \en^j_A(x) = \eta^{ij},
	\eqno(5.13) $$
where $\eta^{ij}$ is the constant metric.
{}From the definition follows that these conformal fields can not be tensor
derived in the sense of (3.13). Namely, $ x_A e^{Ai} \en^i_B = x_B$ so
$ x_A e^{Ai} \ne 0$, and a similar relation holds for the inverse.

For $N$-dimensional conformal fields,
the appropriate frame algebra is $so(N+1)$ (or possibly $so(N,1)$, since the
zeroth direction is special), because only square matrices can have an
inverse. It would also be possible to use
$sl(N+1)$ as frame algebra, but then the metric could not be constant.
Moreover, the description of frame spinors would be unnatural, which is a
problem if applications to physics are a concern. The possible frame algebras
for tensor fields are of course $so(N)$ and $gl(N)$.

Under the frame algebra $Map(N, so(N+1))$ ,
$$ [J^{ij}(m), J^{kl}(n)] = \eta^{ik} J^{jl}(m+n) -  \eta^{il} J^{jk}(m+n)
-  \eta^{jk} J^{il}(m+n) +  \eta^{jl} J^{ik}(m+n),
	\eqno(5.14) $$
the vielbein transforms as a vector,
$$ J^{ij}(m) e^{Ak} = \eta^{jk} e^{Ai} - \eta^{ik} e^{Aj},
	\eqno(5.15) $$
as does $\en^k_A$.

The vielbein and its inverse can be regarded as forms
$e_1^i \in \Omega^1(0,c)$ and $\en_{-1}^i \in \Omega_1(0,c)$.
The corresponding gauge connection
$\omega_1^{ij}$ is usually called spin connection in the tensorial case.
For completeness we write down the conformal analog of Cartan's
structure equations for the torsion $T_2^i$ and the curvature $R_2^{ij}$.
$$ \eqalign{
&T_2^i = D_1(\omega,0,c) e_1^i
= d_1(0,c) e_1^i + \omega_1^{ij} \wedge e_1^j \cr
&R_2^{ij} = D_1(\omega,0,c) \omega_1^{ij}
= d_1(0,c) \omega_1^{ij} + \omega_1^{ik} \wedge \omega_1^{kj} \cr
&D_2(\omega,0,c) T_2^i = R_2^{ij} \wedge e_1^j \cr
&D_2(\omega,0,c) R_2^{ij} = 0
}	\eqno(5.16) $$

The vielbein enables a translation between conformal and frame
indices. If $\phi_A \in \CC^0_1(\lambda,c)$ is a frame scalar,
$\phi^i \equiv e^{Bi} \phi_B \in \CC^0_0(\lambda,c)$ is a frame vector, etc.
Using the exterior derivative on scalar fields, maps from arbitrary
conformal fields can now be constructed, e.g.,
$$ \nabla^A \phi_B = \en^i_B (\d^A + {\lambda \over c} k^A) e^{Ci} \phi_C
+ \en^i_B \omega^{Aij} e^{Cj} \phi_C
	\eqno(5.17) $$
{}From this relation the transformation law for conformal Christoffel symbols
can be extracted, but it is not very illuminating.

\endchapter
\header{6. The zeroth dimension }

{}From the previous section it is clear that conformal fields in $N$
dimensions in many ways behave as tensor fields in $N+1$ dimensions,
simply because of the number of components. One striking effect is the
$so(N+1)$ frame algebra.  It is tempting to take this feature
seriously and interpret the extra zeroth dimension as time.  The frame
algebra would then be a gauged Lorentz symmetry which could act as the
arena for flat space physics.  This possibility is investigated in
this section. However, although our language is colored by this
physical interpretation, which may or may not be correct, the
mathematical results are perfectly sensible.

There is an obvious extension of (3.2) to an algebra of vector fields
in $N+1$ dimensions; simply replace the tensor index $\mu$ by a
conformal index $A$.  However, using the building blocks of (3.3),
there could also be a term proportional to $k^A$ which vanishes when
$A = \mu$. Hence the most general expression reads
$$ L^A(m) = \emx \bigg( \d^A + (m^B + k^B) T^A_B
+ k^A (\alpha m^B + \beta k^B) T^C_B x_C + c m^A m^B T^C_B x_C \bigg),
	\eqno(6.1) $$
where the parameters $\alpha$ and $\beta$ are arbitrary.

The $L^A(m)$ do not generate all of $Vect(N+1)$, which is clear
already from the fact that $m^0$ is not independent of $m^\mu$.  This
is not really physically surprising, because we do not expect the time
component of the momentum to be independent of its space components.
To see what kind of algebra that the $L^A(m)$ generate, we limit
ourselves to the scalar representation $T^A_B = 0$; other
representations yield more complicated expressions.
$$ \eqalign{
[L^A(m), L^B(n)] &= (n^A + k^A) L^B(m+n) - (m^B + k^B) L^A(m+n) \cr
[L^A(m), n^B] &= - k^B n^A \enx \cr
[L^A(m), \enx] &=  n^A \emnx
}	\eqno(6.2) $$
and all other brackets vanish. This algebra is similar
to, but distinct from, $Vect(N+1)$, and our ``time'' is therefore
not a dimension
in the same sense as the $N$ space dimensions.

By setting $m = 0$ in (6.1), we obtain the {\it momentum operator}
$$ P^A_\beta = \d^A + k^B T^A_B + \beta k^A k^B T^C_B x_C.
	\eqno(6.3) $$
The momenta satisfy the same
relations as the derivatives $\d^A$, namely
$$ \eqalign{
[P^A_\beta, P^B_{\beta^\prime}]
&= k^A P^B_{\beta^\bis} - k^B P^A_{\beta^\bis} \cr
[P_\beta^A, x_B] &= \delta^A_B - k^A x_B
} \qquad \eqalign{
[P_\beta^A, m^B] &= - k^B m^A \cr
[P_\beta^B, \emx] &= m^A \emx
}	\eqno(6.4) $$
where ${\beta^\bis}$ is arbitrary.
The {\it Hamiltonian} is
$$ P^0_\beta = - x\ddot\d + \beta T^\sigma_0 x_\sigma + (1+\beta) T^0_0
	\eqno(6.5) $$
It commutes with the $gl(N)$ generators $J^\mu_\nu$ (3.8), but
the remaining bracket $[P^0_\beta, K_\nu]$ is complicated.

A scalar conformal field $\phi(x) \in \CC^0_0(\lambda, c)$ satisfies
$$ P^A_\beta \phi(x) = (\d^A + \lambda (1+\beta) k^A) \phi(x),
	\eqno(6.6) $$
and in the particular case that it is translationally and scale
invariant ($\d^A \phi = 0$), it is a momentum eigenvector. The only
non-zero eigenvalue is $\lambda (1+\beta)$ in the ``time'' direction,
and thus we think of this quantity as the mass. This is closely
related to the dilatation operator
$$ D = {{\d L^\mu(m)} \over {\d m^\mu}} \Bigg|_{m=0}
= x\ddot\d + T^\mu_\mu,
	\eqno(6.7) $$
with eigenvalue is $\lambda$.

The value $\alpha = \beta = -1$ is special, for two related reasons.
The ``mass'' is zero independent of the value of $\lambda$, and
$x_A L^A(m) \equiv 0$. In this case, the following relation holds.
$$ [L^\mu(m), P_{-1}^A] = m^A L^\mu(m) - \emx P_{-1}^A
	\eqno(6.8) $$
Otherwise this bracket is complicated.

\endchapter

\newpage

\noindent\header{References}

\noindent 1. T. A. Larsson, to appear in Int. J. Mod. Phys. A (1992)

\noindent 2. E. Ramos and R. E. Shrock, Int. J. Mod. Phys. A 4 (1989) 4295.

\noindent 3. E. Ramos, C. H. Sah and R. E. Shrock, J. Math. Phys. 31
(1989) 1805.

\noindent 4. F. Figueirido and E. Ramos, Int. J. Mod. Phys. A 5 (1991) 771.

\noindent 5. T. A. Larsson, Phys. Lett. B 231 (1989) 94.

\noindent 6. T. A. Larsson, J. Phys. A 25 (1992) 1177.

\noindent 7. P. Goddard and D. Olive, Int. J. Mod. Phys. A 1 (1986) 303.

\noindent 8. T. Eguchi, P. B. Gilkey and A. J. Hansson, Phys. Rep. 66
(1980) 213.

\noindent 9. C. Nash and S. Sen, Topology and geometry for physicists,
(London: Academic Press, 1983).

\vfill
\end